# Spontaneous Enhancement of Inward Impurity Pinch during Massive Impurity Injection in Tokamak Plasmas

L. Zeng[1,†], W. Xia[2,3,†], D. Hu[4], T. Tang[2], X. Zhu[2], S. Zhao[5], T. Shi[2], D. Chen[2], Y. Duan[2], L. Xu[2], G. Li[2], G. Zuo[2], J. Hu[2], X. Gao[2], Y. Sun[2,*], Z. Gao[1,*]

[1] Department of Engineering Physics, Tsinghua University, Beijing 100084, China
[2] Institute of Plasma Physics, Hefei Institutes of Physical Science, Chinese Academy of Sciences, Hefei 230031, China
[3] University of Science and Technology of China, Hefei 230026, China
[4] School of Physics, Beihang University, Beijing 100191, China
[5] Institute of Energy, Hefei Comprehensive National Science Center, Hefei, 230031, China

† These authors contributed equally to this work.
*Contact authors: ywsun@ipp.ac.cn and gaozhe@tsinghua.edu.cn

A spontaneous enhancement of inward impurity pinch is identified prior to the fast quench under massive impurity injection on the EAST tokamak. The transition is marked by a rapid shift of the magnetic-island rotation frequency toward the electron diamagnetic direction within a few hundred microseconds. An extended KDG model shows that the impurity-induced increase in bulk ion collisionality can drive a transition across banana regime, nonlinearly modifying the parallel viscosity and producing a radial electric field excursion of approximately −10 kV/m, consistent with experiment. The associated self-organized response of the plasma flow and electrostatic potential can modify the parallel friction and further strengthen the neoclassical inward impurity pinch. Projections for ITER disruption mitigation suggest that this mechanism is expected to enhance impurity penetration, thereby improving the overall effectiveness of mitigation.

## Introduction

Plasma–material interaction is a key subject across various fields of plasma physics, astrophysical plasma and lab plasma [1,2]. Material ablation in a hot plasma can provide radiative cooling via atomic physics and lead to large-scale dynamics [3]. In the field of magnetic fusion energy research, detailed studies of material ablation in plasmas could help design methods to prevent serious wall damage from major disruptions in future tokamak-based powerplants [4–6]. A variety of rapid impurity injection methods have been explored to safely dissipate plasma energy, including shattered pellet injection (SPI) [7], massive gas injection (MGI) [8] and dispersive shell pellet injection [9]. Advances have been made in rapid shutdown research and the SPI method has been chosen as the disruption mitigation system for ITER [10,11]. However, the efficiency of these methods in future devices is highly uncertain, partially due to an incomplete understanding of impurity penetration and ablation dynamics.

Depositing impurities in the plasma core is the most effective method for disruption mitigation, but the penetration of injected impurities is limited by several mechanisms during their transport from the plasma surface. Charge exchange and recombination strongly brake the injected impurity through effective frictional drag and an opposing neutral pressure gradient force [12]. The pellet rocket effect, arising from asymmetric plasma heating of the ablation cloud, can significantly reduce the impurity penetration depth [5]. Compounding this limitation is the polarization induced by $\nabla\mathbf{B}$ drift in the presence of high local plasmoid pressure, which can drive impurities toward the low-field side (LFS) [13–15]. This drift favors high-field side (HFS) injection but hinders LFS injection. Although large MHD modes can induce impurity mixing in the core region [3,8,16], it becomes most pronounced only near the end of thermal quench (TQ) [8], and is accompanied by substantial heat flux losses [17,18], too late to overcome the early stage penetration barrier or ensure effective mitigation.

Here, we report experimental evidence from the EAST tokamak showing that inward impurity penetration is spontaneously enhanced during massive impurity injection, accompanied by a clear increase in core plasma radiation. The injection of massive impurities induces a rapid increase in bulk ion collisionality, transitioning the bulk plasma from the banana regime to Pfirsch-Schlüter (P-S) regime. Consequently, this transition triggers a reorganization of the plasma flow and a negative excursion of the radial electric field, while the associated change in parallel friction variation can in turn strengthen the inward impurity convection.

**Experimental results**

Disruptions are deliberately triggered by injection of large amounts of neon or argon via SPI or MGI on EAST, with auxiliary heating provided by neutral beam injection (NBI) or electron cyclotron resonance heating (ECRH). Further details can be found in [19].

Figure 1 shows a disruption triggered by SPI, where approximately $6.7\times10^{21}$ neon particles were injected at t = 0 ms into co-current NBI heated plasmas. The spatiotemporal evolution in the horizontal array of absolute extreme-ultraviolet (AXUV) emission [Fig. 1(c)] exhibits two distinct stages prior to the final fast quench stage, which is characterized by a rapid drop in the core temperature associated with global stochasticity [18,20]. During the earlier stage, the radiation presents a hollow profile, indicating that the impurities are primarily concentrated in the plasma edge. At ~2 ms, the core emission strengthens abruptly while the edge rapidly weakens, implying a sudden penetration of impurities across the $q = 2$ surface. Meanwhile, an intrinsic $m/n = 2/1$ mode undergoes a phase reversal toward the electron diamagnetic direction within a few hundred microseconds, suggesting a strong connection for the two phenomena. The enhanced penetration favors impurity assimilation and radiative dissipation of the core thermal energy, as indicated by the substantial decrease in core temperature following the transition.

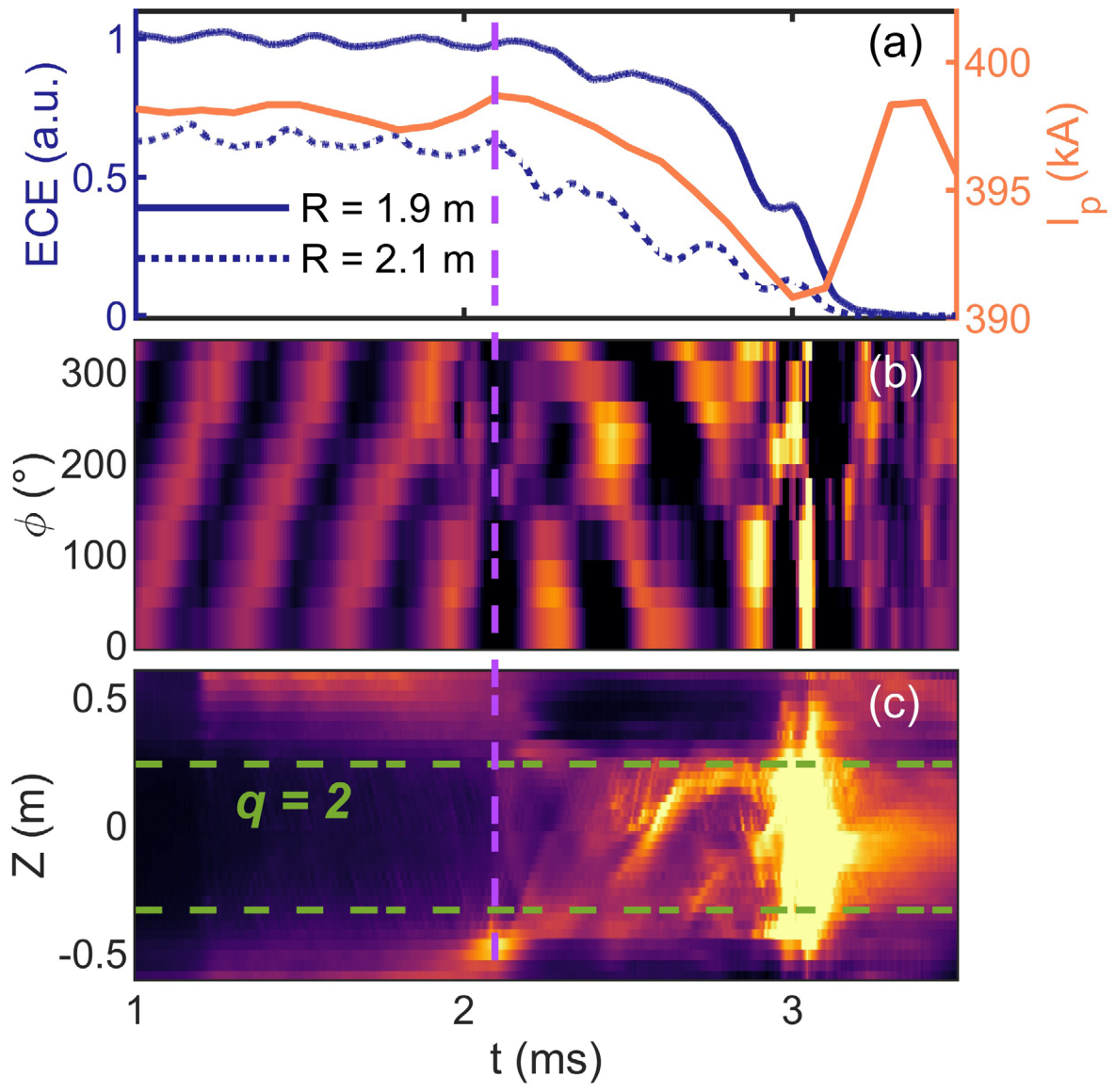

FIG. 1. Time evolution of key diagnostics in the SPI experiment. (a) Electron cyclotron emission (ECE) signals at two radial positions (R = 1.9 and 2.1 m) together with the plasma current $I_p$. (b) Toroidal propagation of the magnetic perturbation. (c) Spatiotemporal evolution of the AXUV radiation perturbation measured along horizontal chords. The green dashed lines mark the $q = 2$ surface, and the purple dashed lines indicate the mode reversal.

Similar results are also confirmed in neon MGI experiments. For a clearer demonstration of the impurity injection trajectory, MGI experimental observations are preferred, as SPI involves the simultaneous injection of multiple impurity fragments. Figs. 2(a) and 2(b) present the vertical and horizontal AXUV measurements, respectively. Initially, the impurity radiation front penetrates inward gradually while drifting poloidally in the ion diamagnetic direction, but is stopped in the vicinity of the plasma edge, consistent with observations on other devices [8,21]. At ~ 6 ms, vertical AXUV measurement shows sharp inward penetration and the soft-X-ray (SXR) signals exhibit a strong profile contraction, indicating an abrupt intensification of core radiation after a characteristic transition, similar to that in Fig. 1(c). By combining measurements from two intersecting SXR arrays [one of which is shown in Fig. 2(c)] with the AXUV measurements [Figs. 2(a), (b)], the location of the intense emission spot associated with the concentrated impurity cloud is reconstructed in Fig. 2(d). Prior to the transition, impurities are largely deposited outside the $q = 2$ surface and propagate at ~60 m/s. Following the transition, the penetration velocity rises sharply to at least ~300 m/s, enabling impurities to cross the $q = 2$ surface.

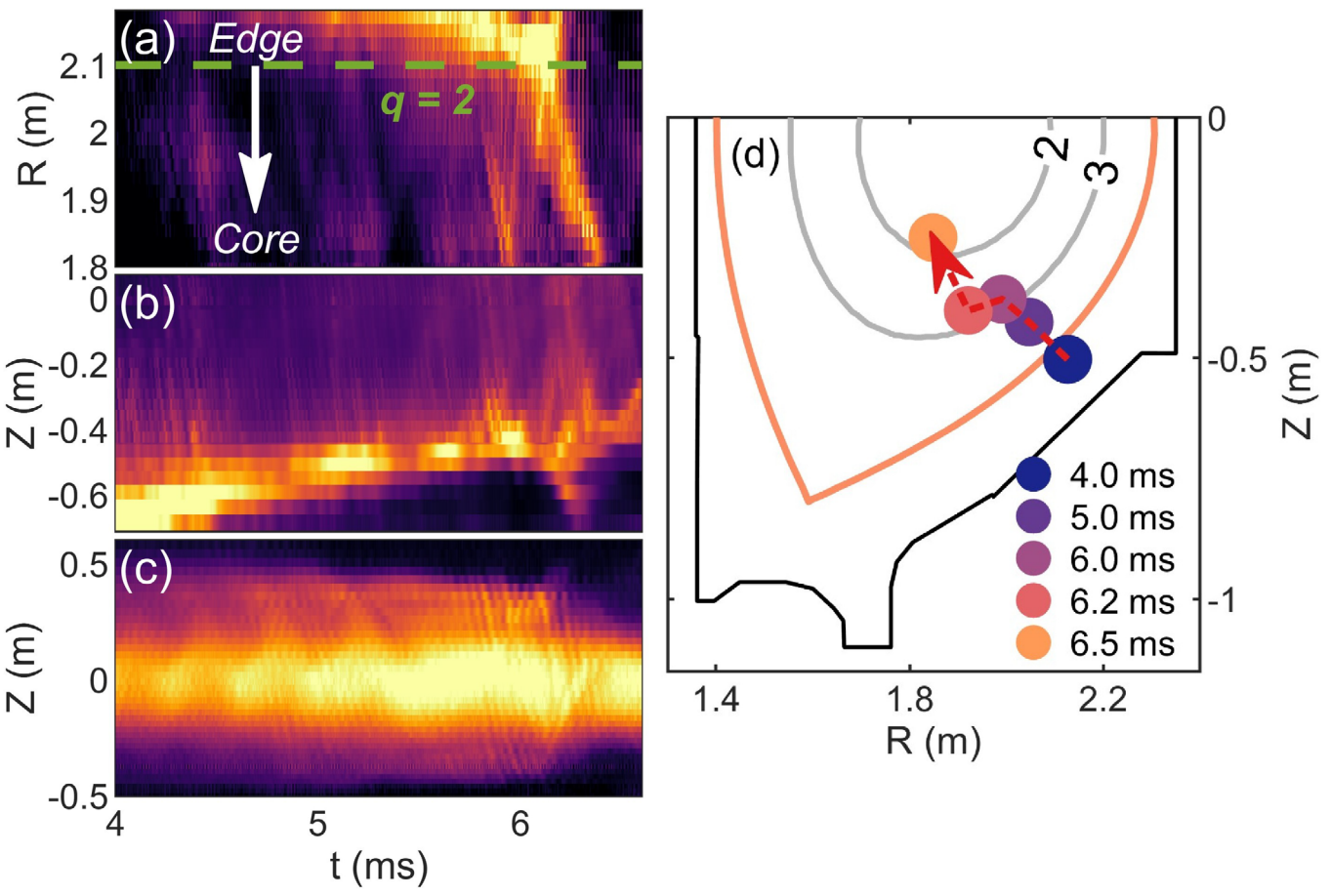

FIG. 2. Spatiotemporal evolution of (a) vertical and (b) horizontal AXUV radiation perturbations, and (c) horizontal SXR emission in the MGI case. (d) Reconstructed inward propagation trajectory of the impurity radiation front, together with the locations of the $q = 2$, 3 surfaces and the last closed flux surface.

Figure 3 presents the magnetic perturbations measured by poloidal and toroidal Mirnov probe arrays. The phase relation between two poloidal positions exhibits a lead–lag reversal at ~5.7 ms [red and blue lines in Fig. 3(b)], and the same reversal is also observed in the toroidal signals [Fig. 3(c)], indicating a reversal in the mode propagation direction. This low-frequency coherent perturbation is identified as an intrinsic rotating $m/n = 2/1$ magnetic island, whose frequency decreases from +2 kHz and then reverses to −3 kHz within 300 μs, where the positive sign is defined in the ion diamagnetic drift direction. The poloidal rotation variation of this mode is consistent with the observed poloidal variation of impurity deposition [Fig. 2(d)]. Furthermore, although the mode amplitude grows continuously after impurity injection [gray dashed line in Fig. 3(b)], the impurity transport is not observed to strengthen progressively [Fig. 3(d)], as would be expected if it were induced by large-amplitude MHD activity [3,22]. Instead, the enhanced penetration drives the mode growth, indicating the penetration enhancement is distinct from the rapid impurity mixing driven by MHD activity.

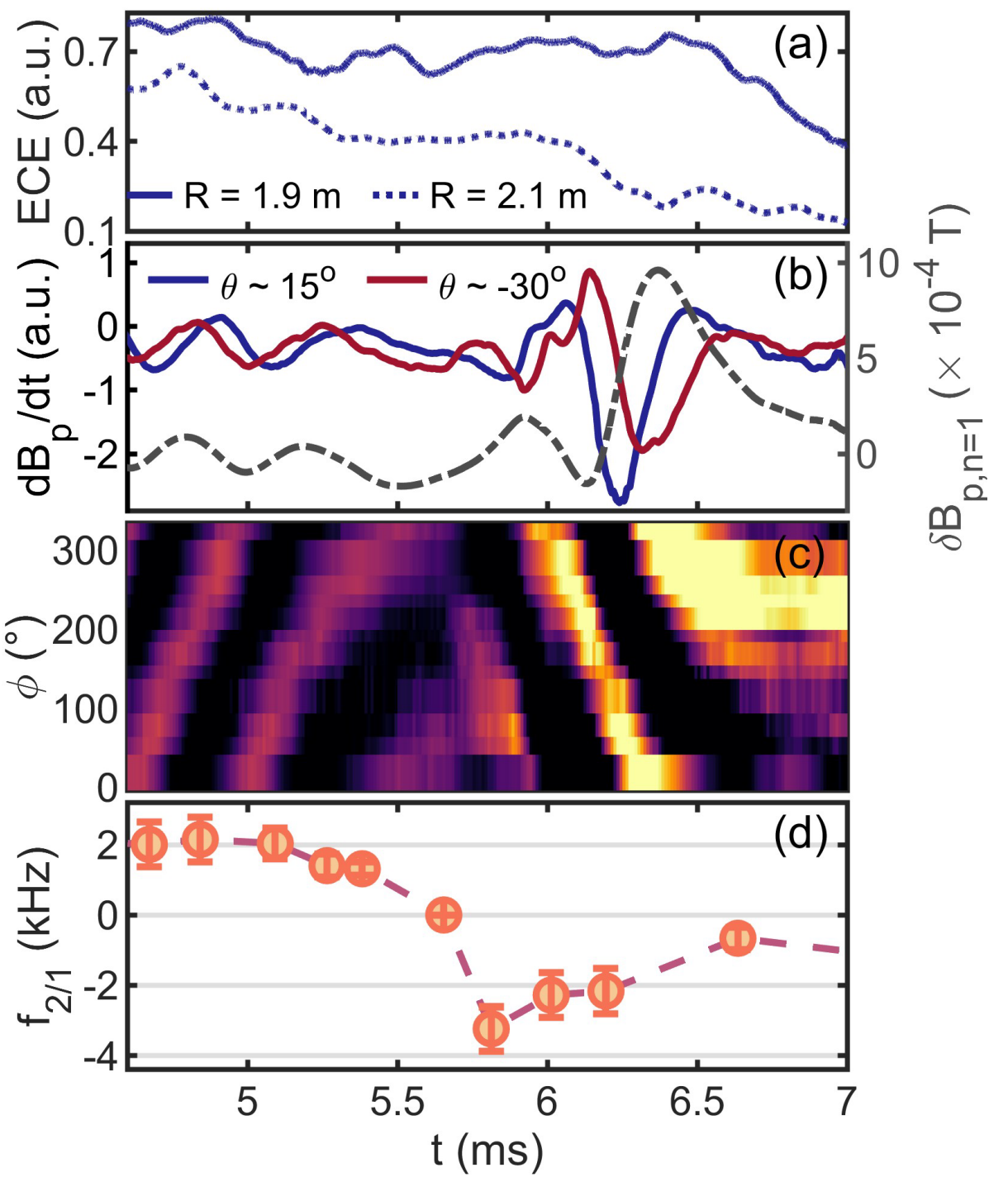

FIG. 3. Evolution of signals in the MGI case. (a) ECE signals at two radial positions (R = 1.9 m and 2.1 m) (b) Differential magnetic perturbation signals measured at two poloidal angles (θ ~ 15° and ~ -30°), showing a clear evolution of the phase lead–lag relation between the two locations. The gray dash-dotted line represents the n = 1 component of the integrated poloidal perturbation measured at θ ~ 15°. (c) Toroidal propagation of the magnetic perturbation. (d) the *m*/*n* = 2/1 mode frequency, with error bars indicating the measurement uncertainty.

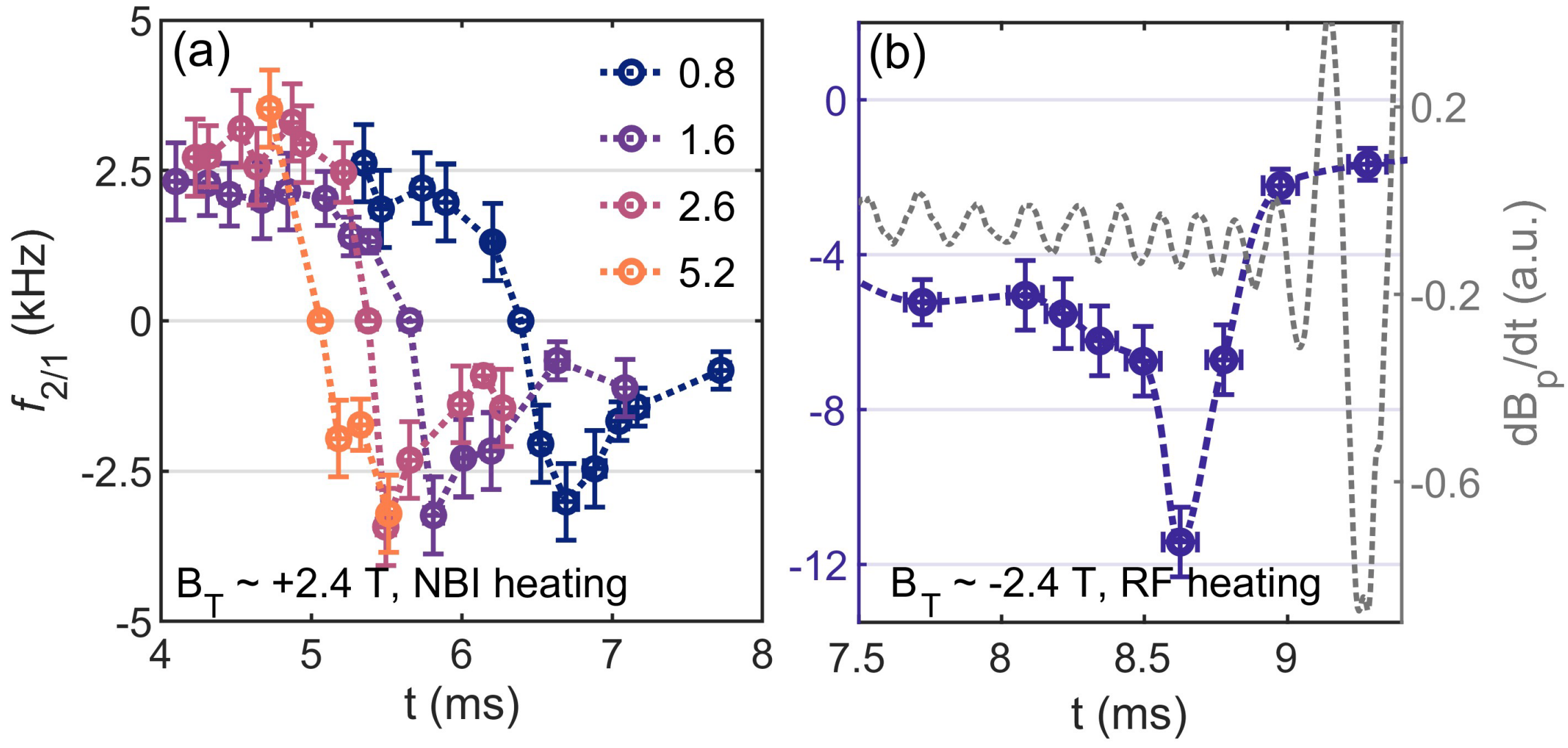

FIG. 4. (a) Evolution of the *m*/*n* = 2/1 mode frequency during neon gas injection with different amounts in NBI-heated plasma. Legend values are in units of $10^{21}$ injected particles. (b) Temporal evolution of the 2/1 mode frequency and magnetic perturbations during argon injection into an RF-heated plasma.

Fig. 4(a) shows how the 2/1 mode frequency evolves with injected impurity amount in MGI cases. In all cases, the frequency shifts from ~ +2.5 kHz to ~ –3 kHz. With increasing injected impurity amount, the transition occurs earlier and becomes more rapid. A similar change in mode frequency is obtained with RF heating, shown in Fig. 4(b), where the frequency shifts directly from about –5 kHz to –11 kHz. Note that NBI can modify the toroidal rotation and thereby produce the initial ion diamagnetic frequency, while the RF heating does not, but both cases reveal a shift toward the electron diamagnetic frequency after impurity injection. These observations present that a general mechanism drives the mode rotation toward the electron diamagnetic direction on several hundred microsecond timescales even with different injected amounts, impurity species, heating methods, and toroidal magnetic field directions.

**Neoclassical effects of massive impurity injection**

The physics of the mode frequency variation is examined using the standard expression of the magnetic island frequency [23,24]:

$$f_{MHD} = \frac{n}{2\pi}\left(\frac{d\Phi}{d\psi} + \frac{1}{n_e e_0}\frac{dP_e}{d\psi} + 0.71\frac{1}{e_0}\frac{dT_e}{d\psi}\right) \tag{1}$$

Here, $n$ is toroidal mode number and $\psi$ is the poloidal magnetic flux. The three terms represent the **E**×**B** frequency, the electron diamagnetic frequency and the frequency associated with the heat force, respectively. During the transition, no clear change in plasma temperature is observed near the $q = 2$ surface [Figs. 1(a) and 3(a)], indicating that the temperature gradient is unlikely to be the dominant driver. The energy confinement even before TQ is still on the timescale of millisecond on EAST, implying that the electron pressure gradient is not the primary driving mechanism. The reversal is therefore primarily attributed to the **E**×**B** contribution, suggesting a negative radial electric field develops rapidly, driving both the island's rotation and plasma poloidal flow toward the electron diamagnetic direction [25].

By attributing the observed change in 2/1 mode's rotation to the **E**×**B** contribution, the change of radial electric field $\Delta E_r \approx 2\pi R B_p \Delta f_E$ is estimated to shift by approximately -13 kV/m (NBI-dominated) and -14 kV/m (RF-dominated), corresponding to the frequency variations $\Delta f_E$ of about -5.5 and -6 kHz, respectively. Since the sudden change in plasma poloidal rotation, neoclassical parallel viscosity variations are expected to play a dominant role [26–28]. For the typical parameters on EAST, the neoclassical poloidal flow relaxation time in the collisionless regime is $\tau_\theta \sim 1.5\tau_{ii}\varepsilon$ [29], approximately 200 μs. This timescale is consistent with the observed dynamics.

Expression for the neoclassical poloidal and toroidal rotation of the bulk ions and the impurity ions are given by the KDG model [30]. However, rapid transition imposes a strong constraint on the relevant flow dynamics. The transition occurs on a timescale of a few hundred microseconds, much shorter than the toroidal rotation relaxation time (on the order of the energy confinement time [31], ~ 10 ms). The toroidal rotation therefore cannot respond appreciably and the bulk ion parallel flow is approximated as frozen, whereas the poloidal flow can undergo significant changes through $E_r$ to satisfy the neoclassical poloidal flow [32,33]. During massive impurity injection, the parallel viscosity of the impurity could play more important effect, so an extended KDG model has been developed, adopting an instantaneous force balance approximation between neoclassical parallel viscosity and friction for one impurity species and one bulk ion species (see Appendix). $E_r$ can be solved self-consistently from the radial force balance.

For the neoclassical parallel viscosity, the asymptotic dimensionless coefficients μ, given by Kim *et al.* [30,34,35], and further corrected by Callen *et al.* [36] by including collisional boundary-layer effects, are employed. These μ primarily depend on the bulk ion collisionality parameter $\nu_{*,i} = (\nu_{ii}+\nu_{iI})R_0q/(\varepsilon^{1.5}v_{T,i})$ [36–39], where the impurity contribution enters through bulk ion-impurity collisions ($\nu_{iI}$). The friction matrix can be found in Ref. [30], and depends on the impurity strength parameter $\alpha \equiv n_I Z_I^2/n_i Z_i^2$. In this letter, neutral collisions are neglected, because JOREK simulations indicate that the neutral fraction near the island is less than 10% [12,18], and the enhanced heat flux induced by the island further promotes ionization near the impurity cloud front [40]. Within this framework, μ can be modified substantially by $\nu_{*,i}$ induced by $\nu_{iI}$, resulting in a nonlinear change of $E_r$ on the bulk ion collision timescale.

The explicit expression for $E_r$ including impurity viscosity is presented in the Appendix. Neglecting impurity viscosity yields the following simplified form for the injection-induced change ($\Delta E_r$), which provides a lower bound on $\Delta E_r$ and captures the dominant physical mechanism:

$$\Delta E_r > \frac{[K_1(t_0) - K_1(t_1)]T_i}{e_i L_{Ti}} \ (V/m) \quad (2)$$

Here $K_1$, defined in the Appendix following Ref. [30], is the ratio of μ whose collisionality dependence provides the primary source of nonlinearity. It decreases nonlinearly from about 0.5 in the banana regime to about -2 in the P-S regime.

Using macroscopic parameters from EAST diagnostics and the impurity-related parameter from JOREK simulation results [18], we adopt representative values of $B_T = -2.4$ T, $B_p = 0.2$ T, $n_i = 4\times10^{19}$ m$^{-3}$, $T_i = 600$ eV, $T_I = 100$ eV, and consider deuterium bulk ions and low-charge-state neon impurities ($Z_I = 3$). The prescribed gradient scale lengths are $L_{pi} = -0.05$ m, $L_{pI} = 0.05$ m, $L_{TI} = -0.05$ m, where $L_X^{-1} = d(\ln X)/dr$. The local safety factor is set to $q = 2$.

Figure 5 presents the calculated response. As $n_I$ increases, $\nu_{*,i}$ rises, allowing the bulk plasma shift out of the banana regime. The corresponding change in parallel viscosity requires a stronger **E**×**B** response to maintain the balance between viscosity and friction, independent of whether the viscosity model of [30] or [36] is used. The full calculation follows the trend implied by Eq. (2), with $E_r$ shifting toward more negative values as α increases and showing little dependence on the toroidal field direction, in agreement with experiment. It also highlights the importance of the bulk ion temperature gradient. As shown in Fig. 5(b), a steeper temperature gradient yields a larger $|\Delta E_r|$. This dependence offers a natural explanation for why enhanced impurity transport is observed near the separatrix of the 2/1 island, where strong and sustained temperature gradients are expected [41]. The nonmonotonic behavior of $\Delta E_r$ near $\alpha \sim 10$ in Fig. 5(b) is attributed to impurity viscosity [30]. In JOREK simulations of MGI in EAST [18], the maximum $\nu_{*,i}$ reaches about 40 while the temperature changes only modestly, extending into the P-S regime, which makes the $\Delta E_r$ predicted by the above mechanism reasonable. Fig. 5(b) displays the estimated $\Delta E_r$ is on the order of -10 kV/m, and the corresponding frequency shift ($\Delta f_E$) of approximately -6 kHz, in good agreement with the experimental value. As the injected impurity amount increases, $\nu_{iI}$ grow more rapidly, indicating an earlier and faster transition process, consistent with Fig. 4(a).

Furthermore, JOREK simulations for ITER have also shown that the maximum $\nu_{*,i}$ can reach about 10 during SPI triggered disruptions [42]. This suggests that the same mechanism should not be neglected under ITER-relevant conditions and may drive a larger $|\Delta E_r|$ because of the higher temperature and steeper temperature gradient. This interpretation is further supported by Ref. [33], where simulations showed that the abrupt establishment of a temperature gradient in a neoclassical plasma can drive a rapid excursion of $E_r$.

Motivated by the predominantly convective nature of the observed penetration and the accompanying neoclassical rotation dynamics, the neoclassical impurity pinch driven by parallel friction is investigated [43–46]. The flux-surface-averaged (FSA) neoclassical radial impurity flux satisfies $Ze\langle \mathbf{\Gamma}_I^{\mathrm{neo}} \cdot \nabla\psi \rangle = -RB_T\langle \mathbf{B} \cdot \mathbf{F}_{I1}/B^2 \rangle$ [45], showing that the flux is determined by parallel friction and magnetic geometry. Because the present FSA treatment does not resolve the poloidal structure of the impurity density or friction force, it cannot provide a quantitative prediction of the radial flux. Nevertheless, $\langle \mathbf{B} \cdot \mathbf{F}_{I1} \rangle$ (calculated as described in Appendix) provides an intuitive indication about impurity transport. Fig. 5(c) shows $\langle \mathbf{B} \cdot \mathbf{F}_{I1} \rangle$ can grow nonlinearly with $\nu_{*,i}$, reflecting the self-organized response of the coupled plasma flow and electrostatic potential system, which further enhances the inward impurity pinch upon crossing the banana regime boundary. Note that the strong poloidal inhomogeneity that can arise during massive impurity injection [8,47] may therefore produce an inward impurity pinch substantially larger than the FSA estimate. These results suggest a positive feedback mechanism for the observed two stage penetration, in which impurity accumulation in the outer region drives the local collisionality across banana regime boundary and thereby strengthens the inward pinch through the self-organized response. Such a mechanism may enhance impurity assimilation and improve disruption mitigation in ITER and other fusion devices.

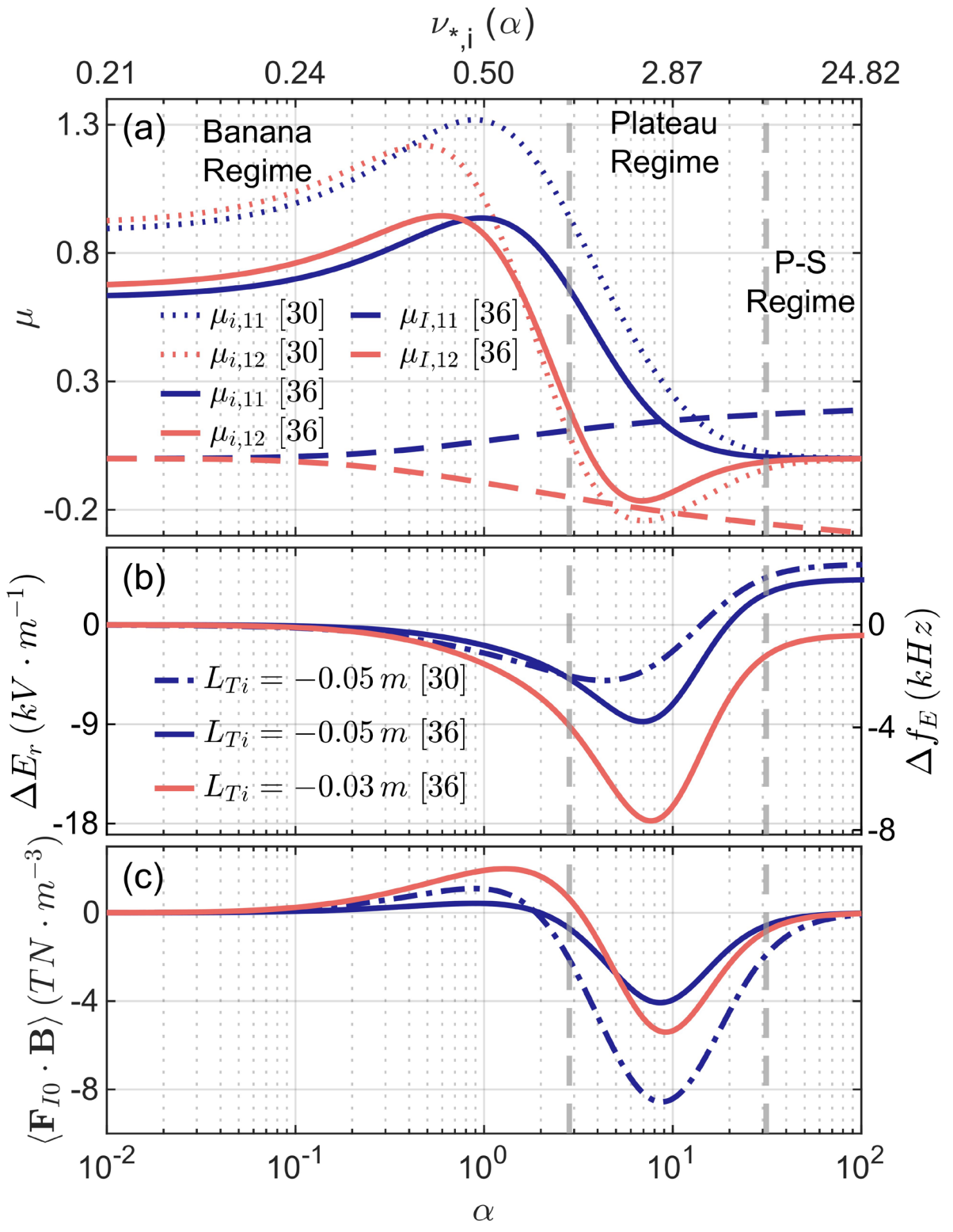

FIG. 5. (a) Asymptotic dimensionless coefficients μ, (b) $\Delta E_r$ (left axis) and the corresponding $\Delta f_E$ (right axis), as functions of the impurity strength α (with the corresponding bulk ion collisionality parameter indicated along the top axis). (c) FSA parallel friction. The gray dashed lines mark the transitions between the banana and plateau regimes and between the plateau and P-S regimes. The viscosity coefficients model used for each curve is indicated by the reference label given in the legend.

**Summary**

An abrupt enhancement in impurity penetration is observed in massive impurity injection experiments on EAST, accompanied by a mode frequency transition toward the electron diamagnetic direction within a few hundred microseconds. The frequency shift implies a rapid excursion of $E_r$ of order -10 kV/m. An extended KDG model is developed, demonstrating that massive impurity injection can drive the bulk plasma collisionality across banana regime to P-S regime through bulk ion–impurity collisions, thereby substantially modifying the bulk ion parallel viscosity and driving $E_r$ toward more negative values. The resulting self-organized readjustment of the plasma flows and electrostatic potential modifies the parallel friction and can thereby strengthen the neoclassical inward impurity pinch. Projections for ITER indicate that the maximum $\nu_{*,i}$ could reach P-S regime during SPI triggered disruptions. Taking into account this effect will therefore enhance impurity penetration, thereby improving the overall effectiveness of mitigation in ITER.

This work was supported by the National Magnetic Confinement Fusion Research and Development Program of China (Grant No. 2024YFE03010000), and the National Natural Science Foundation of China (Grant Nos. 12475225 and 12405269). We thank the staff members at EAST in Hefei (https://cstr.cn/31130.02.EAST), for providing technical support and assistance in data collection and analysis.

[1] G. E. Morfill and A. V. Ivlev, Complex plasmas: An interdisciplinary research field, Rev. Mod. Phys. **81**, 1353 (2009).

[2] J. Deca, A. Divin, P. Henri, A. Eriksson, S. Markidis, V. Olshevsky, and M. Horányi, Electron and Ion Dynamics of the Solar Wind Interaction with a Weakly Outgassing Comet, Phys. Rev. Lett. **118**, 205101 (2017).

[3] D. G. Whyte, T. E. Evans, A. W. Hyatt, T. C. Jernigan, R. L. Lee, A. G. Kellman, P. B. Parks, R. Stockdale, and P. L. Taylor, Rapid Inward Impurity Transport during Impurity Pellet Injection on the DIII-D Tokamak, Phys. Rev. Lett. **81**, 4392 (1998).

[4] C. Reux et al., Demonstration of Safe Termination of Megaampere Relativistic Electron Beams in Tokamaks, Phys. Rev. Lett. **126**, 175001 (2021).

[5] N. J. Guth, O. Vallhagen, P. Helander, I. Pusztai, S. L. Newton, and T. Fülöp, Pellet Rocket Effect in Magnetic Confinement Fusion Plasmas, Phys. Rev. Lett. **134**, 035101 (2025).

[6] A. Matsuyama, R. Sakamoto, R. Yasuhara, H. Funaba, H. Uehara, I. Yamada, T. Kawate, and M. Goto, Enhanced Material Assimilation in a Toroidal Plasma Using Mixed H 2 + Ne Pellet Injection and Implications to ITER, Phys. Rev. Lett. **129**, 255001 (2022).

[7] D. Bonfiglio et al., Validation of 3D MHD simulations of $Ne/D_2$ mixed shattered pellet injection in JET, Nucl. Fusion **66**, 026014 (2026).

[8] E. M. Hollmann et al., Measurements of injected impurity assimilation during massive gas injection experiments in DIII-D, Nucl. Fusion **48**, 115007 (2008).

[9] E. M. Hollmann, P. B. Parks, D. Shiraki, N. Alexander, N. W. Eidietis, C. J. Lasnier, and R. A. Moyer, Demonstration of Tokamak Discharge Shutdown with Shell Pellet Payload Impurity Dispersal, Phys. Rev. Lett. **122**, 065001 (2019).

[10] I. Bandyopadhyay et al., MHD, disruptions and control physics: Chapter 4 of the special issue: on the path to tokamak burning plasma operation, Nucl. Fusion **65**, 103001 (2025).

[11] P. Barabaschi et al., Progress of ITER and its importance for fusion development, Nucl. Fusion (2026).

[12] E. Nardon, A. Fil, P. Chauveau, P. Tamain, R. Guirlet, H. R. Koslowski, M. Lehnen, C. Reux, F. Saint-Laurent, and J. E. T. Contributors, On the mechanisms governing gas penetration into a tokamak plasma during a massive gas injection, Nucl. Fusion **57**, 016027 (2016).

[13] V. Rozhansky, I. Veselova, and S. Voskoboynikov, Evolution and stratification of a plasma cloud surrounding a pellet, Plasma Phys. Control. Fusion **37**, 399 (1995).

[14] P. T. Lang, K. Büchl, M. Kaufmann, R. S. Lang, V. Mertens, H. W. Müller, J. Neuhauser, ASDEX Upgrade Team, and NI Team, High-Efficiency Plasma Refuelling by Pellet Injection from the Magnetic High-Field Side into ASDEX Upgrade, Phys. Rev. Lett. **79**, 1487 (1997).

[15] P. B. Parks and L. R. Baylor, Effect of Parallel Flows and Toroidicity on Cross-Field Transport of Pellet Ablation Matter in Tokamak Plasmas, Phys. Rev. Lett. **94**, 125002 (2005).

[16] D. Hu, E. Nardon, M. Hoelzl, F. Wieschollek, M. Lehnen, G. T. A. Huijsmans, D. C. Van Vugt, S.-H. Kim, Jet Contributors, and the Jorek Team, Radiation asymmetry and MHD destabilization during the thermal quench after impurity shattered pellet injection, Nucl.

Fusion **61**, 026015 (2021).
[17] M. Lehnen et al., Disruption mitigation by massive gas injection in JET, Nucl. Fusion **51**, 123010 (2011).
[18] W. Xia et al., Interpretive JOREK simulation of thermal quench triggered by massive gas injection in EAST disruptions, Nucl. Fusion **65**, 056028 (2025).
[19] X. Gong et al., Overview of recent experimental results on the EAST Tokamak, Nucl. Fusion **64**, 112013 (2024).
[20] ITER Physics Expert Group on Disruptions, Plasma Control, and MHD, Chapter 3: MHD stability, operational limits and disruptions, Nucl. Fusion **39**, 2251 (1999).
[21] R. S. Granetz et al., Gas jet disruption mitigation studies on Alcator C-Mod and DIII-D, Nucl. Fusion **47**, 1086 (2007).
[22] E. M. Hollmann, T. C. Jernigan, E. J. Strait, G. Antar, T. E. Evans, D. S. Gray, M. Groth, D. A. Humphreys, P. B. Parks, and D. G. Whyte, Observation of q-profile dependence in noble gas injection radiative shutdown times in DIII-D, Physics of Plasmas **14**, 012502 (2007).
[23] M. F. Heyn, I. B. Ivanov, S. V. Kasilov, and W. Kernbichler, Kinetic modelling of the interaction of rotating magnetic fields with a radially inhomogeneous plasma, Nucl. Fusion **46**, S159 (2006).
[24] Y. Kikuchi et al., Forced Magnetic Reconnection and Field Penetration of an Externally Applied Rotating Helical Magnetic Field in the TEXTOR Tokamak, Phys. Rev. Lett. **97**, 085003 (2006).
[25] F. L. Waelbroeck, Theory and observations of magnetic islands, Nucl. Fusion **49**, 104025 (2009).
[26] K. Ida and J. E. Rice, Rotation and momentum transport in tokamaks and helical systems, Nucl. Fusion **54**, 045001 (2014).
[27] K. C. Shaing, Test of tokamak low-mode–high-mode transition theory in stellarators, Physics of Fluids B: Plasma Physics **5**, 3841 (1993).
[28] K. C. Shaing and E. C. Crume, Bifurcation theory of poloidal rotation in tokamaks: A model for L-H transition, Phys. Rev. Lett. **63**, 2369 (1989).
[29] F. L. Hinton and M. N. Rosenbluth, Dynamics of axisymmetric and poloidal flows in tokamaks, Plasma Phys. Control. Fusion **41**, A653 (1999).
[30] Y. B. Kim, P. H. Diamond, and R. J. Groebner, Neoclassical poloidal and toroidal rotation in tokamaks, Physics of Fluids B: Plasma Physics **3**, 2050 (1991).
[31] J. H. F. Severo, I. C. Nascimento, V. S. Tsypin, and R. M. O. Galvão, Plasma residual rotation in the TCABR tokamak, Nucl. Fusion **43**, 1047 (2003).
[32] K. C. Shaing, K. Ida, and S. A. Sabbagh, Neoclassical plasma viscosity and transport processes in non-axisymmetric tori, Nucl. Fusion **55**, 125001 (2015).
[33] S. V. Novakovskii, C. S. Liu, R. Z. Sagdeev, and M. N. Rosenbluth, The radial electric field dynamics in the neoclassical plasmas, Physics of Plasmas **4**, 4272 (1997).
[34] S. P. Hirshman and D. J. Sigmar, Neoclassical transport of impurities in tokamak plasmas, Nucl. Fusion **21**, 1079 (1981).
[35] Y. B. Kim, P. H. Diamond, and R. J. Groebner, Erratum: "Neoclassical poloidal and

toroidal rotation in tokamaks" [Phys. Fluids B 3, 2050 (1991)], Phys. Fluids B **4**, 2996 (1992).

[36] J. D. Callen, Viscous Forces Due To Collisional Parallel Stresses For Extended MHD Codes, No. UW-CPTC 09-6R, University of Wisconsin, 2010.

[37] S. Newton and P. Helander, Neoclassical momentum transport in an impure rotating tokamak plasma, Physics of Plasmas **13**, 012505 (2006).

[38] I. Calvo, F. I. Parra, J. L. Velasco, and J. M. García-Regaña, Impact of main ion pressure anisotropy on stellarator impurity transport, Nucl. Fusion **60**, 016035 (2020).

[39] W. Fundamenski and O. E. Garcia, Comparison of Coulomb Collision Rates in the Plasma Physics and Magnetically Confined Fusion Literature, No. EFDA–JET–R(07)01, EFDA-JET, 2007.

[40] W. Li, J. Li, Z. Lin, J. Q. Dong, J. T. Luo, and Y. Liu, Excited ion-scale turbulence by a magnetic island in fusion plasmas, Sci Rep **14**, 25362 (2024).

[41] L. Bardóczi, A. Dudkovskaia, R. J. La Haye, J. D. Callen, C. Chrystal, and M. Podesta, Perturbed Ion Temperature and Toroidal Flow Profile Measurements in Rotating Neoclassical Tearing Mode Magnetic Islands, Phys. Rev. Lett. **132**, 065107 (2024).

[42] D. Hu, E. Nardon, F. J. Artola, M. Lehnen, D. Bonfiglio, M. Hoelzl, G. T. A. Huijsmans, S.-J. Lee, and the JOREK Team, Collisional-radiative simulation of impurity assimilation, radiative collapse and MHD dynamics after ITER shattered pellet injection, Nucl. Fusion **63**, 066008 (2023).

[43] C. Angioni, Impurity transport in tokamak plasmas, theory, modelling and comparison with experiments, Plasma Phys. Control. Fusion **63**, 073001 (2021).

[44] P. Helander, Bifurcated neoclassical particle transport, Physics of Plasmas **5**, 3999 (1998).

[45] C. Angioni and P. Helander, Neoclassical transport of heavy impurities with poloidally asymmetric density distribution in tokamaks, Plasma Phys. Control. Fusion **56**, 124001 (2014).

[46] T. Fülöp and P. Helander, Nonlinear neoclassical transport in toroidal edge plasmas, Phys. Plasmas **8**, 3305 (2001).

[47] M. Lehnen, S. N. Gerasimov, S. Jachmich, H. R. Koslowski, U. Kruezi, G. F. Matthews, J. Mlynar, C. Reux, P. C. De Vries, and JET contributors, Radiation asymmetries during the thermal quench of massive gas injection disruptions in JET, Nucl. Fusion **55**, 123027 (2015).

# End Matter

## Appendix: Extended KDG Model

To interpret the abrupt transition of the poloidal flow observed in the experiment, given that the relevant timescale is comparable to bulk ion collision time, the neoclassical parallel viscosity should be modeled [26–28]. Because the viscous response is coupled to the parallel heat flow moment, the parallel momentum and heat flow balance equations must be solved simultaneously to close the system. An instantaneous force balance approximation is adopted, retaining only the parallel momentum and heat flow balance for one bulk ion species and one impurity species:

$$\langle \mathbf{B} \cdot \nabla \cdot \Pi_a \rangle = \langle \mathbf{B} \cdot \mathbf{F}_{a1} \rangle$$
$$\langle \mathbf{B} \cdot \nabla \cdot \Theta_a \rangle = \langle \mathbf{B} \cdot \mathbf{F}_{a2} \rangle$$

Here, $\Pi_a$ and $\Theta_a$ are the viscous stress tensor and the heat viscous stress tensor, respectively, while $\mathbf{F}_{a1}$ and $\mathbf{F}_{a2}$ are the friction and the heat friction force, respectively.

Assuming $x = v_{t,I}/v_{t,i} \ll 1$, the equations can be written in the matrix form:

$$\frac{n_I m_I \tau_{ii}}{n_i m_i \tau_{II}} \begin{bmatrix} \mu_{11}^I & \mu_{12}^I \\ \mu_{21}^I & \mu_{22}^I \end{bmatrix} \begin{bmatrix} u_{1\theta}^I \langle B^2 \rangle \\ u_{2\theta}^I \langle B^2 \rangle \end{bmatrix} = -\begin{bmatrix} \alpha & 0 \\ 0 & \alpha \frac{T_i}{T_I}\left(7.5 + \sqrt{2}\alpha\, x^{-1}\right) \end{bmatrix} \begin{bmatrix} \langle u_{1\parallel}^I B \rangle \\ \langle u_{2\parallel}^I B \rangle \end{bmatrix} + \begin{bmatrix} \alpha & 1.5\alpha \\ 0 & 6.75\alpha x^2 \end{bmatrix} \begin{bmatrix} \langle u_{1\parallel}^i B \rangle \\ \langle u_{2\parallel}^i B \rangle \end{bmatrix}$$

$$\begin{bmatrix} \mu_{11}^i & \mu_{12}^i \\ \mu_{21}^i & \mu_{22}^i \end{bmatrix} \begin{bmatrix} u_{1\theta}^i \langle B^2 \rangle \\ u_{2\theta}^i \langle B^2 \rangle \end{bmatrix} = -\begin{bmatrix} \alpha & 1.5\alpha \\ 1.5\alpha & \sqrt{2} + 3.25\alpha \end{bmatrix} \begin{bmatrix} \langle u_{1\parallel}^i B \rangle \\ \langle u_{2\parallel}^i B \rangle \end{bmatrix} + \begin{bmatrix} \alpha & 0 \\ 1.5\alpha & 6.75 \frac{T_i}{T_I} \alpha\, x^2 \end{bmatrix} \begin{bmatrix} \langle u_{1\parallel}^I B \rangle \\ \langle u_{2\parallel}^I B \rangle \end{bmatrix}$$

where $\alpha \equiv n_I Z_I^2 / n_i Z_i^2$. Since the transition timescale is much shorter than the toroidal rotation relaxation time, we adopt a timescale separation. Neglecting the θ dependence and taking $B_t^2/\langle B^2 \rangle \simeq 1$, the parallel flow is approximated by the toroidal flow, so that the bulk ion parallel flow $u_{0\parallel}^i$ is effectively frozen during the transition. The radial force balance and the heat flow satisfy the following relations:

$$\begin{bmatrix} E_r \\ \langle u_{2\parallel}^i B \rangle \end{bmatrix} = \begin{bmatrix} -\frac{B_p}{B_t} u_{1\theta}^i \langle B^2 \rangle \\ u_{2\theta}^i \langle B^2 \rangle \end{bmatrix} + \begin{bmatrix} \frac{B_p}{B_t} \langle u_{1\parallel}^i B \rangle + \frac{T_i}{e_i} L_{pi}^{-1} \\ \frac{B_T T_i}{B_p e_i} L_{T_i}^{-1} \end{bmatrix}$$

$$\begin{bmatrix} \langle u_{1\parallel}^I B \rangle \\ \langle u_{2\parallel}^I B \rangle \end{bmatrix} = \begin{bmatrix} u_{1\theta}^I \langle B^2 \rangle \\ u_{2\theta}^I \langle B^2 \rangle \end{bmatrix} + \begin{bmatrix} -\frac{B_T T_I}{e_I}\left(\frac{L_{pI}^{-1}}{B_p} - \frac{e_I E_r}{T_I B_p}\right) \\ \frac{B_T T_I}{B_p e_I} L_{T_I}^{-1} \end{bmatrix}$$

Collecting terms yields the full equation set:

$$\begin{bmatrix} \left(\frac{n_I m_I \tau_{ii}}{n_i m_i \tau_{II}}\right)\mu_{11}^I + \alpha & \left(\frac{n_I m_I \tau_{ii}}{n_i m_i \tau_{II}}\right)\mu_{12}^I & \frac{\alpha B_T}{B_p} & -1.5\alpha \\ \left(\frac{n_I m_I \tau_{ii}}{n_i m_i \tau_{II}}\right)\mu_{21}^I & \left(\frac{n_I m_I \tau_{ii}}{n_i m_i \tau_{II}}\right)\mu_{22}^I + \alpha\frac{T_i}{T_I}\left(7.5 + \sqrt{2}\alpha x^{-1}\right) & 0 & -6.75\alpha x^2 \\ -\alpha & 0 & -\frac{\mu_{11}^i B_T}{B_p} - \frac{\alpha B_T}{B_p} & \mu_{12}^i + 1.5\alpha \\ -1.5\alpha & -6.75\left(\frac{T_i}{T_I}\right)\alpha x^2 & -\frac{\mu_{21}^i B_T}{B_p} - \frac{1.5\alpha B_T}{B_p} & \mu_{22}^i + \left(\sqrt{2} + 3.25\alpha\right) \end{bmatrix} \begin{bmatrix} u_{1\theta}^I B^2 \\ u_{2\theta}^I B^2 \\ E_r \\ u_{2\theta}^i B^2 \end{bmatrix}$$

$$= \begin{bmatrix} \alpha\frac{B_T T_I}{e_I B_p L_{pI}} + \alpha U_{1\parallel}^i B + \frac{1.5\alpha B_T T_i}{B_p e_i L_{Ti}} \\ -\frac{\alpha\frac{T_i}{T_I}\left(7.5 + \sqrt{2}\alpha x^{-1}\right) B_T T_I}{B_p e_I L_{TI}} + \frac{6.75\alpha x^2 B_T T_i}{B_p e_i L_{Ti}} \\ -\mu_{11}^i U_{1\parallel}^i B - \frac{\mu_{11}^i B_T T_i}{e_i B_p L_{pi}} - \alpha U_{1\parallel}^i B - \frac{1.5\alpha B_T T_i}{B_p e_i L_{Ti}} - \frac{\alpha B_T T_I}{B_p e_I L_{pI}} \\ -\mu_{21}^i U_{1\parallel}^i B - \frac{\mu_{21}^i B_T T_i}{B_p e_i L_{pi}} - 1.5\alpha U_{1\parallel}^i B - \frac{\left(\sqrt{2} + 3.25\alpha\right) B_T T_i}{B_p e_i L_{Ti}} - \frac{1.5\alpha B_T T_I}{e_I B_p L_{pI}} + \frac{6.75\frac{T_i}{T_I}\alpha x^2 B_T T_I}{B_p e_I L_{TI}} \end{bmatrix}$$

This full coupled equation set is solved for the impurity poloidal flow $u_{1\theta}^I$, impurity poloidal heat flow $u_{2\theta}^I$, radial electric field $E_r$ and bulk ion poloidal heat flow $u_{2\theta}^i$. In this formulation, the bulk ion poloidal flow can adjust rapidly through the change in $E_r$ on the bulk ion collision timescale.

If impurity viscosity is neglected, the expression reduces to Eq. (2) with a clearer physical interpretation, where the explicit definition of $K_1$ adopted in this work is provided by

$$K_1 = \frac{\mu_{12}^i\left(\sqrt{2} + \alpha - \alpha\beta\right)}{\mu_{11}^i\left(\mu_{22}^i + \sqrt{2} + \alpha - \alpha\beta\right) - \left(\mu_{12}^i\right)^2}$$

Here, $\beta$ is defined as

$$\beta = \left(\frac{27}{4}\right)^2\left(\frac{m_i}{m_I}\right)^2\left(\frac{15}{2} + \sqrt{2}\alpha\frac{v_{Ti}}{v_{TI}}\right)^{-1}$$